\documentclass[allclo]{FBSart}
\usepackage{amsfonts}
\usepackage{amssymb}

\def\Journal#1#2#3#4{{#1} {\bf#2} (#3) #4}

\def\PRL{Phys. Rev. Lett.}

\def\PRC{{Phys. Rev.} C}

\title{Elastic Form Factors of $^{3,4}$He up to Large $Q^2$}
\author{Kees de Jager\thanks{\textit{E-mail address:} 
kees@jlab.org}\\
for the JLab Hall A Collaboration}
\institute{Jefferson Laboratory, Newport News, VA 23606, USA}

\runningauthor{Kees de Jager}
\runningtitle{Elastic Form Factors of $^{3,4}$He up to Large $Q^2$}
\begin{document}

\maketitle
\begin{abstract}
Elastic electron scattering off $^3$He and $^4$He has recently been studied at forward and backward scattering angles in Hall A at JLab. The results will provide accurate data on the elastic form factors, charge and magnetic for $^3$He and charge only for $^4$He, up to squared momentum transfer $Q^2$-values of 3.2 GeV$^2$.
\end{abstract}

\section{Introduction}
A major focus of the research program at the Jefferson Laboratory (JLab) is to test the applicability of hadronic degrees of freedom in few-body nuclear systems through elastic electron scattering and photo-disintegration. Earlier measurements have shown that a hadronic description works well down to scales of 1 fm at low energy transfers. However, deuteron photo-disintegration results show evidence of the scaling predicted by quark counting rules and a failure of the hadronic description at large energy and momentum transfers. The experiment described here aims at an additional test on the $^3$He and $^4$He nuclei.

Unpolarized $ed$ scattering has been studied in Hall A\cite{alexa} at JLab up to a $Q^2$ value of 6 GeV$^2$, providing accurate data on the deuteron structure function $A$. Tensor polarization in that same reaction was studied in Hall C\cite{abbott} up to 1.7 GeV$^2$, allowing a separate determination of all three deuteron form factors $G_E$, $G_M$ and $G_Q$ in that $Q^2$-range. Theoretical calculations using relativistic hadronic theory were able to describe all experimental data up to the highest $Q^2$-value. On the other hand, the data for $A$ follow the scaling behavior predicted by pQCD for $Q^2$-values larger than 4 GeV$^2$.

Photo-disintegration of the deuteron has also been extensively studied at JLab, with publications of cross-section measurements\cite{bochna} from all three Halls and a variety of polarization studies\cite{wije} in Hall A. Hadronic theories can only describe the photo-disintegration cross-section data at low photon energies, while the data show the onset of approximate scaling for transverse momentum $p_T$ values larger than 1.3 GeV/$c$. The prediction of the quark-gluon string (QGS) model also agrees well with the data. The lowest-energy polarization-transfer data agree well with hadronic calculations. Above 500 MeV the magnitude of the induced polarization decreases rapidly to zero, in agreement with hadron helicity conservation (HHC). The polarization transfer coefficients show maxima near 1 GeV and appear to steadily decrease above that energy. This behavior is also consistent with HHC.

At larger energies there are significant differences between the kinematics of $ed$ elastic scattering and deuteron photo-disintegration. In elastic scattering the invariant mass $W$ remains equal to the mass of the deuteron, whereas in photo-disintegration $W = \sqrt{2 \omega m_d + m_d^2}$ and resonances are explicitly excited. For a photon energy of 4 GeV over 200 resonances can contribute through intermediate states, making detailed microscopic hadronic calculations unfeasible.

\section{Theory}

In the one-photon exchange approximation the cross section for elastic scattering off $^3$He is given by:

\begin{equation} 
\frac{d\sigma}{d\Omega} = \frac{ \sigma_M}{\epsilon}[\mu^2 \tau F_M^2(Q^2) + \epsilon F_E^2(Q^2) ] (\frac{1}{1 + \tau}),
\end{equation}

with $M$ and $\mu$ the mass and magnetic moment of the target nucleus, $\tau = Q^2 /(4 M^2)$, $\sigma_{M}=(\frac{Z \alpha_{QED} \cos \theta_e /2}{2 E_e \sin ^2 \theta_e/2})^2 \frac{E_e'}{E_e}$ the Mott cross section and $\epsilon = 1/[1 + 2(1 + \tau) \tan^2(\frac{\theta_e}{2})]$ the linear polarization of the virtual photon. Cross-section measurements at one $Q^2$ value, but different scattering angles, are needed for a separate determination of the charge and magnetic form factors. For the spinless $^4$He nucleus the cross section has only a contribution from the charge form factor

\begin{equation} 
\frac{d\sigma}{d\Omega} =  \sigma_M F_E^2(Q^2) (\frac{1}{1 + \tau}).
\end{equation}

The form factors of the few-body systems are determined by the nuclear charge and current operators, that are a combination of one- and many-body terms. The one-body operator can be expressed in a standard way in terms of single-nucleon convection and magnetization currents. The latter are obtained by calculating the nuclear ground state, for which a number of methods have been used, $e.g.$ solving the Faddeev equations, using correlated hyperspherical harmonics, and Monte Carlo methods, either variational or Green's Function. Modern calculations then include relativistic boost corrections. The two-body current operator can be decomposed in "model-dependent" and "model-independent" components. The model-independent one is obtained from the two-nucleon potential, and is dominated by a pion-like current acting at long range, but also includes a short-range rho-like current. Three-body currents have been estimated, albeit only at small $Q^2$ values, to have only a small contribution. The model-dependent currents are purely transverse and can not be linked to the two-nucleon interaction. The most important contribution, associated with the $\Delta$-isobar, has been shown to have a significant contribution at intermediate $Q^2$ values of $\sim$ 1 GeV$^2$. In the case of the deuteron the calculations are significantly simplified since only isoscalar terms need be included. A recent calculation\cite{schiavil} that also includes a contribution from the $\rho \pi \gamma$ transition mechanism, describes the JLab data\cite{alexa} through the largest $Q^2$ value of 6 GeV$^2$. Most modern calculations are reasonably successful in describing both the charge and magnetic form-factor data for $^3$He and $^4$He through the diffraction maximum but deviate significantly in the region where a second diffraction minimum is expected.

\section{Experiment}

Experiment E04-018\cite{camsonne} was carried out - with two interruptions to run other experiments - from November 2006 through July 2007. An intense electron beam with a current of up to 100 $\mu$A and an energy between 0.7 and 4.4 GeV was incident on cryogenic helium and hydrogen (for calibration measurements) target systems. The helium target cells were 20 cm long and operated at 8 K and 13 atm. The maximum luminosity used in the experiment was $2.10^{38}$ nuclei/cm$^2$/s, which allowed measurements of cross sections as small as $10^{-41}$ cm$^2$/sr at 1 coincident event per day. Scattered electrons and recoiling nuclei were detected in coincidence in the two identical high-resolution spectrometers (HRS). Both HRSs used two planes of scintillators for triggering and timing purposes and two sets of vertical drift chambers for track reconstruction. Electrons were identified using a \v{C}erenkov detector and an electromagnetic lead-glass calorimeter. Electron-nucleus coincidences were identified by double-arm time-of-flight measurements. Data have been obtained in a $Q^2$ range of 1 to 2.6 GeV$^2$ for $^3$He and of 0.4 to 3 GeV$^2$ for $^4$He. The on-line results show a clear indication of a second minimum in the $^4$He charge form factor. If this is corroborated by the full analysis, expected by the summer of 2008, it would be strong evidence that the $^4$He charge form factor is still dominated by hadronic degrees of freedom through $Q^2 \approx$ 2 GeV$^2$.

\begin{acknowledge}
The author thanks the E04-018 collaboration and especially its spokespersons, A. Camsonne, J. Gomez, A.T. Katramatou and G.G. Petratos. This work was supported by DOE contract DE-AC05-06OR23177, under which Jefferson Science Associates, LLC, operates the Thomas Jefferson National Accelerator Facility. 
\end{acknowledge}


\begin{thebibliography}{9}

\bibitem{alexa} Alexa, L.C.,  et al.: \Journal{\PRL} {82} {1999} {1374} 

\bibitem{abbott} Abbott, D., et al.: \Journal{\PRL} {84} {2000} {5053} 

\bibitem{bochna} Bochna, C., et al.: \Journal{\PRL} {81} {1998} {4576}; Schulte, E., et al.: \Journal{\PRL} {87} {2001} {10232}; Rossi, P., et al.: \Journal{\PRL} {94} {2005} {0122301}

\bibitem{wije} Wijesooriya, K., et al.: \Journal{\PRL} {86} {2001} {2975}

\bibitem{schiavil} Schiavilla, R., Pandharipande, V.R.: \Journal{\PRC} {65} {2002} {064009}

\bibitem{camsonne} Camsonne, A., Gomez, J., Katramatou, A.T., Petratos, G.G.: JLab experiment E04-018 

\end{thebibliography}
\end{document}